# Advancing Trustworthy AI in Healthcare Through Meta-Research: Results of an Interdisciplinary Design-Thinking Workshop

Valerie Bürger *[1], Marlie Besouw[2], Jana Fehr[1], Riana Minocher[1], Emma Moorhead[1], Isabel Velarde[1], Louis Agha-Mir-Salim[3,4], Julia Amann[5], Alexandra Bannach-Brown[1,6], David B. Blumenthal[7], Kaitlyn Hair[8], Bert Heinrichs[9], Moritz Herrmann[10,11], Elizabeth Hofvenschiöld[12], Sune Holm[13], Anne A.H. de Hond[14], Sara Kijewski[15], Stuart McLennan[16], Timo Minssen[17], Marco S. Nobile[18], Nico Pfeifer[19], Jessica L. Rohmann[1], Tony Ross-Hellauer[20], Marija Slavkovik[21], Karin Tafur[22], Eleonora Viganò[23], Magnus Westerlund[24], Tracey Weissgerber[1,25] & Vince I. Madai[1,26]

[1] QUEST Center for Responsible Research, Berlin Institute of Health (BIH) at Charité – Universitätsmedizin Berlin, Berlin, Germany
[2] Department of Medical Imaging, Radboud University Medical Centre, Nijmegen, The Netherlands
[3] Institute of Medical Informatics, Charité – Universitätsmedizin Berlin, Corporate Member of Freie Universität Berlin and Humboldt-Universität zu Berlin, Berlin, Germany
[4] Einstein Center Digital Future, Berlin, Germany
[5] Strategy & Innovation, Careum Foundation, Zurich, Switzerland
[6] Institute of Social and Preventative Medicine, University of Bern, Bern Switzerland
[7] Biomedical Network Science Lab, Department Artificial Intelligence in Biomedical Engineering, Friedrich-Alexander-Universität Erlangen-Nürnberg
[8] EPPI Centre, Social Research Institute, University College London, United Kingdom
[9] Institute of Neuroscience and Medicine: Brain and Behaviour (INM-7), Forschungszentrum Jülich, Jülich, Germany; Institute of Science and Ethics (IWE), University of Bonn, Bonn, Germany
[10] Munich Center for Machine Learning (MCML), Munich, Germany
[11] Institute for Medical Information Processing, Biometry, and Epidemiology, Faculty of Medicine, LMU Munich, Munich, Germany
[12] ESB Business School, Reutlingen University, Germany
[13] Department of Food and Resource Economics, University of Copenhagen, Frederiksberg, Denmark
[14] Julius Centre for Health Sciences and Primary Care, University Medical Centre Utrecht, Utrecht University, Utrecht, Netherlands
[15] Department of Health Sciences and Technology, Chair of Bioethics, ETH Zurich, Zurich, Switzerland
[16] Institute of History and Ethics in Medicine, Department of Preclinical Medicine, TUM School of Medicine and Health, Technical University of Munich
[17] Centre for Advanced Studies in Bioscience Innovation Law (CeBIL), Faculty of Law, University of Copenhagen, Denmark
[18] Department of Environmental Sciences, Informatics and Statistics, Ca' Foscari University of Venice, Venice, Italy
[19] Institute for Bioinformatics and Medical Informatics, University of Tübingen, Tübingen, Germany
[20] Open and Reproducible Research Group, Know Center Research GmbH, Graz, Austria
[21] Department of Information Science and Media Studies, University of Bergen, Bergen, Norway
[22] Independent AI Researcher (AI Law and Ethics); Founder and Principal AI Advisor in AI Governance (Policy and Regulation), Paris, France
[23] Digital Society Initiative, University of Zurich; Swiss Institute for Entrepreneurship, University of Applied Sciences of the Grisons, Chur, Switzerland
[24] Department of Business and Economics, and Law, Abo Academy University, Finland; Arcada University of Applied Sciences, Finland
[25] CNC-UC, Center for Neuroscience and Cell Biology, University of Coimbra, Coimbra, Portugal; CIBB, Center for Innovative Biomedicine and Biotechnology, University of Coimbra, Coimbra, Portugal
[26] Faculty of Computing, Engineering and the Built Environment, School of Computing and Digital Technology, Birmingham City University, Birmingham, United Kingdom

*Corresponding author:

Name: Valerie Bürger
Affiliation: QUEST Center for Responsible Research, Berlin Institute of Health (BIH), Charité Universitätsmedizin Berlin, Berlin, Germany
E-mail: valerie.buerger@bih-charite.de

## Abstract

Meta-research and Trustworthy AI (TAI) share common goals, namely improving evidence, robustness, and transparency, yet there is very little interplay between the two fields. To investigate the potential benefits of closer collaboration between the domains of TAI in healthcare and meta-research, we convened an interdisciplinary workshop funded by the Volkswagen Foundation in February 2025. The workshop aimed to collaboratively examine key challenges in translating AI ethics principles into practice and to identify potential solutions informed by meta-research approaches. A Design Thinking-informed co-creation approach was followed by an inductive descriptive analysis of the outputs. Our results demonstrate how meta-research can offer concrete contributions to address pressing challenges of TAI in healthcare. These challenges include the dynamic and complex nature of TAI ethical requirements and principles, common terminology and understanding of TAI, ensuring robustness, replicability, and reproducibility, choosing adequate evaluation metrics, lack of transparency, advancing preclinical biomedical research, and validation in real-world clinical environments. We present a catalog of ideas and a roadmap for future research, which synthesize existing interconnections and identify concrete next steps and open research gaps, thereby serving as a foundation for future interdisciplinary efforts.

## Introduction

Meta-research, also referred to as meta-science or the science of science, aims to describe, explain, evaluate and/or improve scientific practices[1]. As a rapidly evolving field, meta-research in its earliest form largely entailed systematic reviews and synthesis of existing scientific findings[2,3] but has since developed into a broad field that employs theoretical[4,5], experimental[6], and observational[7,8] approaches. It aims to understand, analyze, and improve how science is conducted, ultimately shaping the way scientific knowledge is produced and used[9]. Every stage of the research lifecycle can be subject to analysis and evaluation through meta-research: the composition and behavior of the research community, the choice of research topic, the design and methodology of research studies, the description and interpretation of results, the publication and dissemination of results and the assessment of researchers and funding decisions[10]. Core objectives of meta-research are to analyze and enhance the *transparency* and *robustness* of scientific practice by, for example, identifying the causes of irreproducibility[11,12], establishing standardized reporting guidelines[13,14], analyzing incentive structures that shape scientific behavior[15], and supporting training of research staff[16].

Interestingly, the same concepts – *transparency* and *robustness* – serve as core principles of the field of Trustworthy AI (TAI)[17,18], which focuses on principles and methods that support the ethical development of AI. Within TAI, *transparency* requires that key development information, such as the characteristics of training data, be made openly available[19–21]. *Robustness* refers to the reliable performance of AI systems across diverse settings and patient populations[22,23]. The evident overlap of goals and principles shared between TAI and meta-research points towards the value of examining their intersection.

Partly, this can be explained by the close connection of AI development in healthcare and research. Academic research is not a parallel activity to AI product development but is woven into every stage of it. Higgins and Madai (2020) proposed a four-phase framework, FORM, BUILD, LAUNCH, and MAINTAIN, that maps the lifecycle of an AI medical device from initial concept through to post-market surveillance. As illustrated in Figure 1, each phase of the development is accompanied by a distinct body of research activity that both informs and validates progress. In the FORM phase, literature and scoping reviews establish the evidence base and identify unmet clinical needs and research gaps, while proof-of-concept studies and early publications lay the groundwork for what is technically and clinically feasible. As development transitions into the BUILD phase, research activity intensifies: pilot clinical validation studies, including usability and feasibility assessments, generate the empirical evidence needed to refine the AI system, while data collection and external validation studies extend its generalisability beyond the original development context. The LAUNCH phase demands the most rigorous research standards, with randomised controlled trials and regulatory studies providing the level of evidence required by health authorities to provide evidence that the product is safe and effective and therefore ready for clinical implementation, alongside impact assessments that evaluate real-world clinical benefits and risks. Finally, the MAINTAIN phase recognises that an AI medical device does not exist in a static environment: surveillance studies, ongoing real-world evidence collection, and living systematic reviews ensure that the product continues to perform safely and equitably as clinical practice and patient populations evolve. While the specific development and research process may vary depending on the specific tool and underlying claims being made[24], we use this framework to show that academic research is the scientific backbone of AI development in healthcare. Academic research practices provide the methodology, evidence standards, and critical scrutiny that distinguish trustworthy AI-driven medical products from unvalidated or premature tools.

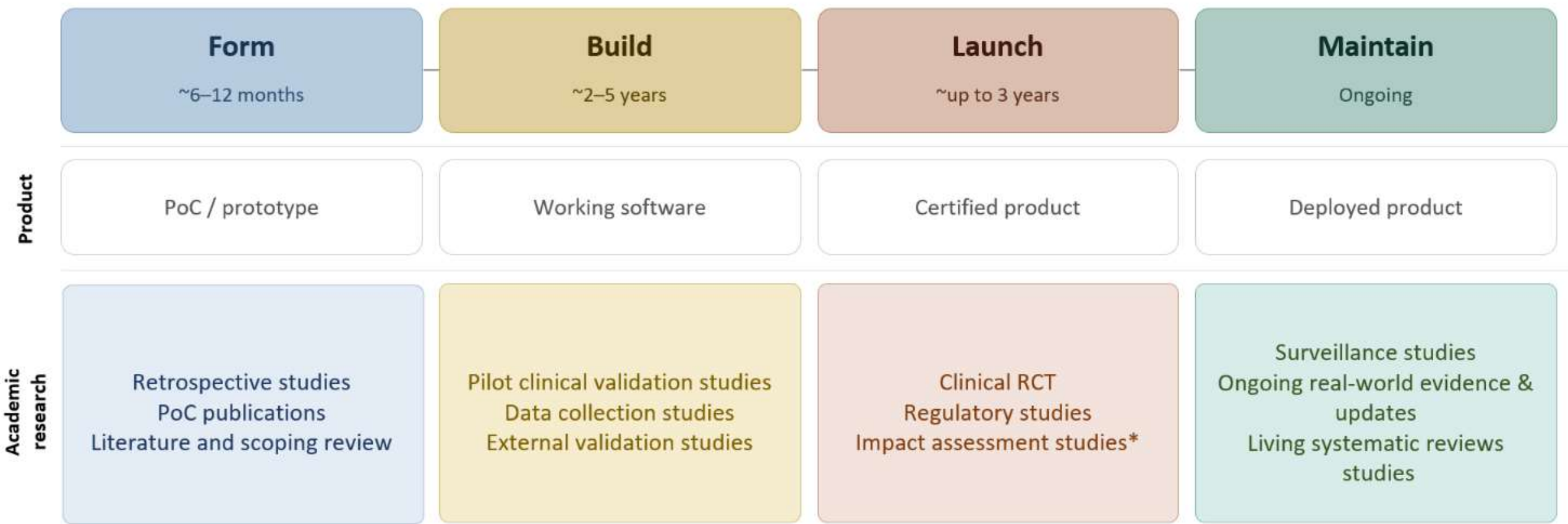

*Figure 1. The figure maps academic research activities across the four phases of AI medical device development (Higgins & Madai, 2020). PoC = Proof of Concept. Pilot clinical validation studies encompass usability, feasibility, and UX studies. RCTs should include an adequate comparator; specific RCT trial design including superiority vs. non-inferiority/equivalence depends on the individual use case. *Impact assessment in the Launch phase refers to clinical impact assessment of the deployed product, not early-stage feasibility assessment.*

Given the central role of research across the entire development lifecycle, meta-research could be an essential element in addressing challenges faced by TAI in healthcare[25,26].Healthcare is a pressing area of concern, given that numerous AI applications are classified as “high risk” under the EU AI Act[27], making the trustworthiness of medical AI tools a critical priority. To date, TAI is already an interdisciplinary field[28], but has been primarily shaped by disciplines such as computer science, ethics, and law[29,30]. Meta-research could therefore fill a critical gap: traditionally focused on medicine; it offers a well-established foundation that could be extended to address the specific challenges TAI faces in healthcare. While reciprocal insights are certainly valuable, this paper examines the potential of meta-research to inform TAI. We explored whether - and in what ways - meta-research could enhance the development, evaluation, and implementation of TAI by fostering greater transparency and robustness.

To answer this question, we convened an interdisciplinary workshop funded by the Volkswagen Foundation[1], which brought together experts from both TAI and meta-research. Through a

[1] https://www.volkswagenstiftung.de/en

Design Thinking–informed co-creation approach[31,32], we collected researchers' perspectives on the interrelatedness of their fields and explored how meta-research could offer concrete contributions to address pressing challenges in TAI in healthcare. We present a descriptive analysis of the collected insights, seven main TAI challenges and corresponding meta-research contributions. There is a particular focus on the dimensions of robustness and transparency, as those principles represent core principles shared by TAI and meta-research. Building on this analysis, we present a catalog of ideas for meta-research activities that can support TAI in overcoming its own challenges and a roadmap for research that prioritizes a sequence of executable actions. While our discussion focuses on AI in healthcare, the lessons drawn here are in many cases applicable across other fields.

The intended audience of this work is primarily researchers and scholars in meta-research and Trustworthy AI, as well as funders, journal editors, and policymakers with an interest in improving scientific practice in medical AI and making AI in healthcare trustworthy. The regulatory and institutional examples drawn upon reflect predominantly European and North American contexts.

## Methods

The analysis presented in this paper is based on materials collaboratively generated at a three-day workshop, *From Principles to Practice: Innovating Trustworthy AI through Meta-research*, funded by the Volkswagen Foundation. The workshop took place in person in Hanover, Germany between February 19th and 21st, 2025. The workshop was conducted in English. This study was not prospectively registered.

*Participants.* Using purposive sampling to engage a diverse spectrum of relevant experts, we recruited 29 researchers with expertise in TAI and/or meta-research. Invitations were extended to researchers with established track records in either TAI, meta-research, or both, particularly those focused on healthcare applications. The selection process ensured diversity in both their academic backgrounds and level of seniority. Beyond TAI and/or meta-research, participants brought further interdisciplinary perspectives from ethics (including medical ethics, neuroethics, and ethics operationalization), computer science, public policy, law, and epidemiology. Participants also encompassed a wide range of academic career stages, including doctoral candidates, postdoctoral fellows, associate professors, full professors, and lead principal investigators. Yet, the selection was skewed towards seniority since this was a requirement by the funder of the workshop. It included 16 women and 13 men and no non-binary or agender person (N = 29). For detailed information on participant demographics, see the Appendix.
The target group size of approximately 30 was chosen to enable meaningful small-group co-creation (six groups of five) while remaining logistically feasible for a three-day in-person event. Participants were recruited via direct personal invitation by the organising committee, drawing on professional networks in both fields. No centralised open call was used, and panelists were not asked to suggest additional members. All participants were reimbursed for travel and accommodation costs by the funding body. No members of the public, patients, or carers were involved in the workshop, as the exercise targeted disciplinary researchers rather than healthcare end-users. The workshop was conducted in English only; no translation or plain-language adaptations were provided

*Approach and Facilitation.* The workshop's approach was based on co-creation, design thinking, and innovation methods[31–33]. A co-creation workshop format was chosen in preference to alternative approaches such as a Delphi exercise or literature review, because the central aim was to generate novel interdisciplinary connections rather than to aggregate pre-existing expert opinions or synthesise documented evidence; the workshop format enabled iterative, dialogic knowledge production that neither a sequential survey method nor a desk-based synthesis could replicate. Unlike traditional approaches where authors write in isolation, this method engaged all stakeholders in an iterative, user-centric process. It emphasized a collaborative process for generating written materials through interactive sessions, as well as fostering cross-disciplinary exchange to encourage collaboration across diverse fields and perspectives[34]. Participants were able to critically analyze and creatively refine the written materials throughout the workshop.

The workshop was facilitated by a professional facilitator with expertise in design thinking as well as a professor of innovation and inclusion at Kristiania University College in Oslo, Norway.

| Pre-Workshop | During Workshop | | | Post-Workshop |
|---|---|---|---|---|
| | Day 1 | Day 2 | Day 3 | |
| Pre-workshop survey | Collaborative SWOT mapping of challenges | Collaborative text generation using guided templates | Peer review of drafted materials | Post-workshop survey |
| | Stakeholder exercise on robustness and transparency | Dialogue sessions for cross-disciplinary exchange | Agreement on continued collaboration and contribution | Content analysis of collaborative writing materials |
| | Group reflection and formulation of overarching questions | Group construction of "ideal" healthcare AI system | | |
| | | Group writing to synthesize workshop insights | | |

Legend
Generation of Text Material
Brainstorming/ Design Thinking Activities

*Figure 2. Overview of workshop activities, including pre-workshop preparation, day-by-day sessions, and post-workshop work. Sessions that contributed to the materials analyzed in this paper are shown in blue.*

*Workshop Agenda.* Before the workshop, participants completed a 24-item survey assessing strengths, challenges, areas of overlap, and opportunities related to TAI and meta-research. The item receiving the most votes for the most significant challenge in TAI for healthcare was "*Operationalizing high-level Trustworthy AI principles, such as robustness and transparency,*" followed by *"Finding consensus Trustworthy AI requirements between stakeholders."* For overlaps between TAI and meta-research in healthcare, *"Transparency as a shared value across both fields"* received the highest number of votes. The complete survey results are provided in the Supplementary Materials.

The results of the pre-work survey guided the activities on the first day. The first day focused on framing discussions and eliciting initial perspectives. Activities included structured introductions, an icebreaker exploring assumptions regarding AI, collaborative SWOT mapping of challenges across both fields, and a stakeholder role-play exercise examining robustness and transparency from multiple viewpoints. The day concluded with group reflection and the formulation of overarching questions to guide subsequent sessions.

The second day centered on solution building and co-creation of workshop outputs. Participants engaged in collaborative text generation using guided templates, short timed "speed collaboration" sessions, and collaborated to construct an "ideal" transparent and robust healthcare AI system. These activities were selected to promote cross-disciplinary exchange and identify actionable steps and ongoing barriers. Participants engaged in group writing to develop a shared set of materials synthesizing insights from the workshop.

The third day focused on refinement and planning of the next steps. Participants conducted peer review of the drafted content, followed by a "commitment wall" activity in which each person outlined concrete post-workshop contributions for finalizing and disseminating the materials. The workshop concluded with closing reflections and agreement on the next steps for continued collaboration. A post-workshop survey was administered to collect participants' feedback and reflections. The complete survey results are provided in the Supplementary Materials.

*Data Generation and Collection*. Upon meeting, participants were divided into six groups, with each group intentionally composed to reflect diversity in research backgrounds and career stages. For the data generation and collection underlying this paper, we conducted a structured, template-based collaborative writing exercise in which participants completed guided prompts on key challenges in operationalizing Trustworthy AI in healthcare and potential meta-research solutions. The guided writing templates were reviewed and refined by the organising committee prior to the workshop but were not formally piloted with external individuals. The workshop was inherently non-anonymous, given its in-person format; participants were aware of each other's identities throughout all sessions. The pre-workshop survey was conducted anonymously via an online platform. The organising committee members participated in workshop activities alongside panelists and contributed to the collaborative writing outputs; their contributions were therefore included in the analysed materials rather than treated as steering input only. The groups contributed heterogeneous material in a shared Microsoft Word document, ranging from brief bullet points and conceptual notes to short narrative passages, documenting their thoughts and discussions, and continuously refining them. At the end of Day 2, each group's draft outputs were circulated to all participants for review and comment; this inter-group feedback was qualitative in nature, delivered through written annotations in the shared document, and was not anonymised. An overview of all of the workshop activities, including those that contributed to the materials analyzed in this paper, is provided in Figure 2. No audio recordings were made at any point in the process.

*Data Analysis.* While the study overall was not designed as a qualitative study, qualitative methods were applied to analyze the generated material. A descriptive analysis[35] was conducted to examine participants' perspectives on the intersection of TAI in healthcare and meta-research. In carrying out extractions and analyzing the material, our goal was to remain close to the original data while inductively identifying recurring patterns and themes[36,37]. The data were read multiple times to ensure familiarity, and initial coding was performed by author VB using the Microsoft Word comment function. An inductive approach was applied. Authors VB and IV collaboratively refined the emerging themes through repeated discussions, revisiting the data, and iteratively developing thematic maps. Sections of the data were re-coded as needed to improve consistency and coherence. Analysis continued until thematic saturation was reached, meaning that no new themes or insights emerged.

## Results

The following section presents seven key TAI issues identified in the workshop, along with their corresponding meta-research contributions. The analysis identified seven key challenges of Trustworthy AI in healthcare and corresponding meta-research contributions. There were three overarching categories emerging from participants' discussions of the workshop materials: (i) overarching challenges in Trustworthy AI, namely the dynamic and complex nature of ethical requirements and the lack of common terminology and understanding; (ii) challenges in the research process for AI in healthcare, including ensuring robustness, replicability, and reproducibility, choosing adequate evaluation metrics, and the lack of transparency; and (iii) challenges specific to particular domains of AI research in healthcare, covering the advancement of preclinical biomedical research and validation in real-world clinical environments.

## Overarching Challenges in TAI in Healthcare

### Dynamic and Complex Nature of TAI ethical requirements and principles

*Key TAI Challenge.* Participants reflected on the *Ethics Guidelines for Trustworthy AI* developed by the European Commission's High-Level Expert Group[17]. These guidelines outline a shared vision of what TAI should achieve. Participants noted that TAI is a multifactorial concept whose requirements can conflict, making its realization a process of continual trade-offs and shifting priorities. They also emphasized that, especially in healthcare, trustworthiness is dynamic and changes over time as technologies, evaluation methods, and social expectations evolve.

*Proposed Meta-research Contribution.* According to participants, meta-research can play an important role in uncovering and describing these tensions in the TAI field by offering a structured approach to their examination. Specifically, meta-research can help to systematically identify and analyze how research on AI is influenced by potential conflicts of principles. Beyond merely mapping these tensions, meta-research also provides a lens to study how researchers acknowledge, negotiate, and, at times, partially resolve such conflicts when designing, implementing, and reporting AI studies. Finally, meta-research could offer tools or conceptual frameworks that help researchers and stakeholders explicitly recognize and deliberate on tensions, rather than resolving them implicitly or ad hoc.

### Common Terminology and Understanding of TAI

*Key TAI Challenge.* Participants identified the lack of common understanding as a significant barrier to implementing TAI. They pointed out a persistent lack of shared understanding of key ethical concepts. One specific example discussed was the concept of reliability. While widely regarded as essential to TAI, it was noted that the term is used inconsistently; sometimes while referring to an AI system's performance stability, in other cases, describing its role in improving broader processes, such as reducing human error. This lack of clarity, they argued, undermines knowledge accumulation and hinders ethical implementation in practice. They observed that when concepts are poorly defined or inconsistently applied, it becomes impossible to meet even minimal ethical standards in practice. Participants warned that lack of a common, operational understanding of ethical principles and related concepts hinders transdisciplinary collaboration and prevents the establishment of a shared conceptual foundation among researchers and experts in both domains.

*Proposed Meta-research Contribution.* Participants recognized the role of meta-research in addressing conceptual fragmentation, particularly through *systematic reviews* that compile and synthesize existing definitions and principles. Such work was viewed as essential for mapping the intellectual terrain of TAI and identifying unresolved tensions or opportunities. However, participants also acknowledged the limitations of synthesis in isolation. They suggested that original conceptual development, grounded in diverse theoretical paradigms, should complement synthesis efforts to meaningfully advance the field's foundational understanding. This would require deliberate efforts to clarify terminology and bridge epistemic divides.

Furthermore, participants called for close transdisciplinary collaboration and the establishment of a shared conceptual foundation among researchers and experts of both domains. Participants proposed that meta-research could eventually serve as a nexus, bringing together diverse research disciplines under a shared umbrella. Fostering close collaboration through meta-research could enable researchers to move beyond traditional silos and more effectively address the complex challenges involved in ensuring the trustworthiness of medical AI tools. However, participants also identified a significant barrier to this vision: the mismatch in pace between the rapid development of AI technologies and the comparatively slower nature of meta-research. They noted that, to remain relevant and impactful in medical-AI contexts, meta-research must become more pragmatic and agile. To help close this gap, participants suggested the development and use of purpose-built AI tools for meta-research, such as automated bias detection systems and platforms for automated systematic reviews, to increase the efficiency and responsiveness of meta-research processes.

## Challenges in the Research Process for TAI in Healthcare

### Ensuring Robustness, Replicability, and Reproducibility

*Key TAI Challenge.* Participants emphasized that ensuring robustness, replicability, and reproducibility in AI development and clinical implementation is a central mission of TAI. These concepts were considered both in the context of technical AI development and the design of AI studies. In addition, some participants explored related distinctions between reproducibility and replicability in more detail: They described replicability as reaching the same scientific conclusions as a previous study, even when using different data or methods, and reproducibility as the ability to duplicate computational results using the same methods, code, data, and software environment. Participants emphasized that reproducibility in AI research and development is fundamental for identifying the sources of variability, distinguishing those arising from data from those introduced by software tools. They further noted that achieving reproducibility depends on transparent documentation and fully repeatable implementation of computational processes, which are often missing in practice. Secure data enclaves, dynamic datasets, and distributed systems, such as those used in federated learning, were described as additional barriers to both reproducibility and replicability. Participants observed that the field lacks a systematic overview of how researchers manage these challenges.

*Proposed Meta-research Contribution.* Meta-research can systematically evaluate how replication studies are conducted, reported, and disseminated, thereby strengthening standards for transparency and comparability across settings. Participants suggested that meta-research could play a key role beyond assessing reporting completeness. In their view, it could address domain-specific questions, such as how to manage and foster replicability in studies and reproducibility in technical implementation, and develop assessment frameworks aligned with these standards. Drawing on experiences from other disciplines facing similar challenges, participants argued that meta-research could both evaluate adherence to existing guidelines but also elevate overall reporting standards and tailor guidelines to address the mentioned challenges, to better capture the complexities of AI research, ultimately strengthening reproducibility and replicability across the field. By examining publication practices, incentive structures, and methodological quality, meta-research can also identify and address barriers, such as the prevailing emphasis on novelty, that limit the visibility and recognition of replication work. Beyond this, meta-research can use approaches such as systematic reviews to generate a comprehensive overview of late-stage tools that have been evaluated in empirical studies across diverse contexts, populations, and implementation settings. By synthesizing and comparing these results, it provides a foundation for assessing the tools' robustness, consistency, and sensitivity to contextual variation.

### Choosing Adequate Evaluation Metrics

*Key TAI Challenge.* One of the key challenges identified by participants in the implementation of TAI in healthcare is the selection of appropriate evaluation metrics. Participants noted the broad recognition in the AI research field that no single metric can fully capture the performance of an AI system: The choice of appropriate evaluation metrics depends not only on the AI method employed but also on the nature of the clinical task, whether it involves binary classification (e.g. cancer detection or cancer prognosis), ranking (e.g. triaging patients for evaluation), or summarization of text (e.g. condensing patient records relevant to a condition). According to participants, this diversity introduces a fundamental challenge for evaluating AI tools: identifying metrics that are fit for purpose, context-specific, and understandable to different stakeholder groups. They further observed that AI systems in healthcare are rarely evaluated based on their ultimate clinical utility – namely, whether implementation results in meaningful improvements in patient outcomes When the goal is to assess impact or effectiveness, predictive performance metrics such as calibration or discrimination are insufficient. Instead, comparative impact studies evaluating real-world outcomes should be used.

*Proposed Meta-research Contribution.* Evaluation metrics must be aligned with the specific question being asked, as predictive accuracy, clinical utility, and impact necessitate different forms of measurement. Once the purpose is identified, as a potential solution to the challenge of metric selection, participants proposed developing a standardized evaluation framework to guide the assessment of AI tools in healthcare. They pointed to initiatives like *Metrics Reloaded*[38] as important steps in this direction, as this initiative supports the appropriate selection of evaluation metrics in biomedical image analysis and provides guidance on aligning metrics with specific tasks and intended use cases. Metrics Reloaded is not, in itself, a meta-research tool; however, it can serve as a foundational framework upon which evaluations of AI studies may be built. In this way, it provides structure and guidance that researchers can draw upon when assessing the quality and reporting practices of AI-focused investigations. Meta-research can establish reporting standards that require explicit specification of the evaluation question, the intended clinical pathway, and patient-centered endpoints, helping to shift the focus from predictive performance alone toward impact-oriented assessment. Building on this, systematic reviews and evidence mapping can quantify the extent to which AI tools lack formal impact evaluations, empirically highlighting gaps in real-world effectiveness evidence.

### Lack of Transparency

*Key TAI Challenge.* Participants described a major transparency gap limiting the ability for oversight and the ability to establish an evidence base to judge the trustworthiness of AI systems. This transparency gap is due to limited information on which AI systems are available on the market, where they are deployed in healthcare, their capabilities, risks, and limitations, as well as their performance and potential shifts in healthcare delivery. Multiple reasons contribute to the lack of transparency in medical AI studies. Participants identified vague regulations, conflicting stakeholder interests, and an idealistic research culture as major barriers to transparency in medical AI. While Article 13 of the EU AI Act requires disclosure of system characteristics, performance metrics, and intended use, its conditional phrasing allows broad interpretation and it does not mandate sharing developmental details with researchers or clinicians, leaving public transparency largely voluntary. Participants furthermore pointed to conflicting stakeholder interests as a key reason for the lack of transparency, for instance companies may view transparency as a competitive risk, while clinicians and patients need sufficient information to weigh benefits and risks but also face concerns about data sharing. Lastly, participants noted that the lack of transparency could be a consequence of a research culture that discourages open discussion of errors. They cautioned against equating trustworthiness with an absence of errors, stressing instead that confidence grows when mistakes are openly documented and corrected rather than concealed.

*Proposed Meta-research Contribution.* Participants emphasized that improving transparency in TAI in healthcare must begin with a clear understanding of the current landscape of AI tools and research. They therefore stressed the need to assess the status quo of transparency, particularly within specific clinical domains. To support this assessment, participants outlined several meta-research-based questions and methods: scoping reviews help by cataloguing existing studies (e.g., how many RCTs exist in a specific domain) and identifying areas that warrant deeper investigation, making estimations about the maturity of a domain possible. Systematic reviews go further by quantitatively assessing how often responsible research practices, like data or code sharing, are reported. These reviews provide detailed snapshots of transparency levels in specific subfields, but only when they themselves follow rigorous, transparent protocols (e.g., PRISMA[13]). Structured paper audits or questionnaires were also proposed as a lighter, less validated, alternative, provided that their selection criteria and process are reproducible. Beyond assessing transparency, participants highlighted promoting proactive practices to foster transparency throughout the AI research pipeline. Key examples included pre-registration of study protocols, for model development, validation studies, or trials, on platforms like OSF, PROSPERO, or clinicaltrials.gov. This not only signals accountability and planning but also ensures public visibility of research intentions and methods, even if amendments are made later. Participants also underscored the importance of reporting guidelines. These are designed to standardize the minimum information disclosed in scientific studies. Promotion of guidelines like TRIPOD+AI[39], TRIPOD-LLM[40], and MINIMAR[41] help ensure consistency in reporting, enabling comparability and preliminary quality assessment of AI systems. However, participants cautioned that while valuable, these guidelines are primarily geared toward expert audiences. For broader public transparency, the same information must be translated into accessible formats and plain language.

As another meta-research tool, participants pointed to registries for approved AI tools, such as the Health AI Register, EUDAMED, or the FDA's AI-Enabled Medical Devices List, as analogous to publication databases. Although these platforms are not always complete and take great effort to maintain, they remain necessary to make informed choices about AI tool implementation. These repositories play a critical role in centralizing information and enhancing visibility into which AI systems are approved, and under what conditions. Lastly, the participants suggested that this issue warrants a meta-research inquiry into how researchers acknowledge and address perceived weaknesses in their systems, arguing that such evidence could inform more effective governance and accountability mechanisms to foster transparency regarding errors and strengthen reproducibility research.

## Challenges in Specific Domains of TAI in Healthcare

### Advancing Preclinical Biomedical Research

*Key TAI Challenge.* Although discussions of TAI often focus on clinical applications, participants emphasized that similar challenges arise when applying AI in preclinical domains, such as drug discovery and molecular biology. In these areas, AI models are primarily used for hypothesis generation and research guidance. Participants stressed that core TAI principles, like technical robustness, transparency, and sound data governance, are equally relevant but frequently unmet in these contexts. Because in-silico results guide costly, animal-intensive wet-lab experiments, participants warned that unstable models for hypothesis generation risk wasting resources and delaying therapeutic development. Single-cell biology was cited as a prominent example. Participants noted that researchers rarely justify their choice of analytical tools when building computational preclinical pipelines, and ad hoc decisions at each step can alter resulting cell labels, undermining robustness and transparency. They pointed out that even small changes in analysis pipelines can lead to substantially different results, raising ethical concerns about using unreliable AI outputs to guide expensive laboratory work involving animals and public funds. Participants further noted that this problem is exacerbated by the, often superficial, treatment of computational methods in the literature: Methods sections commonly reduce complex analytical pipelines to brief statements. Without the release of raw data, well-documented analysis code, and details on preprocessing steps, parameter settings, and model selection, independent laboratories cannot reproduce the analyses. Participants questioned the robustness and reliability of such results and, consequently, the validity of the biological conclusions derived from them.

*Proposed Meta-research Contribution.* Participants outlined a series of concrete, meta-research-informed actions aimed at advancing TAI in preclinical biomedical research: They recommended systematically using meta-research to identify and map evidence gaps, thereby guiding research priorities, and argued that efforts should shift from primarily developing new models toward establishing rigorous benchmarking practices that enable fair comparison and robust evaluation. In addition, they proposed the integration of established scientific practices, such as preregistration of studies and the use of meta-analysis, to enhance credibility and reduce bias of preclinical AI research. Beyond methodological considerations, participants emphasized the importance of adopting and transparently reporting professional software engineering standards, including unit testing, version control, and continuous integration, as a means of improving both reproducibility and transparency in computational workflows. Finally, they underscored the necessity of achieving full reproducibility even before publication, calling for public access to underlying data and code, and suggesting that peer reviewers should actively verify executable pipelines as part of manuscript evaluation process to improve the reliability of reported findings.

### Validation in Real-World Clinical Environments

*Key TAI Challenge.* Participants identified the absence of timely, context-specific late-stage evidence as a major barrier to the trustworthiness and safety of healthcare AI. Late-stage evidence involves assessing an AI model through validation studies in the real-world settings where it is intended to be used, after it has been initially developed and trained. They emphasized that testing mature AI systems in varying real-world clinical contexts is essential for assessing contextual appropriateness, model utility, and ensuring stable performance. Ongoing performance monitoring was viewed as equally critical for detecting tool degradation or failure and enabling timely updates or decommission. Despite this, participants observed that clinical validation, monitoring, and maintenance studies remain limited. Participants consistently emphasized this gap and underscored the need for rigorous validation, drawing analogies to drug development: ideas start in the “lab,” go through phased testing for safety and effectiveness, and continue under post-market surveillance once released. In their view, AI systems should undergo comparable phased testing for safety and effectiveness, followed by continuous post-market surveillance. One participant explained that pharmacological interventions are subject to preclinical testing and randomized controlled trials before approval for use and argued that AI in healthcare should undergo comparably rigorous evaluation. This perspective parallels the impact assessment phase of the clinical risk prediction model lifecycle, in which the central question is not whether a model predicts well, but whether its implementation favorably changes clinical decisions (clinical usefulness) and improves patient outcomes in practice – a question that requires evaluation metrics beyond internal predictive performance.

*Proposed Meta-research Contribution.* Participants regarded meta-research as a central instrument for guiding late-stage validation and monitoring of AI in healthcare. They proposed a "grading" tool tailored for AI research as a meta-research method, which could be used to evaluate the certainty of evidence and strength of implementation recommendations, akin to the GRADE framework used in clinical evidence evaluation[42]*, allowing judgments about the "maturity" of a tool.* A complementary "traffic light" system could visually signal readiness for use. Participants furthermore underscored the value of systematic reviews in consolidating and critically assessing the evidence base for AI tools. Such reviews can evaluate the adequacy of training datasets, scrutinize methodological rigor, and assess whether validation efforts replicate realistic clinical scenarios, and examine whether study designs permit credible causal inferences about clinical impact. They suggested that tools commonly used in systematic reviews, such as "risk of bias" instruments, could be adapted to AI evaluation to assess how methodological choices influence confidence in reported outcomes. Moreover, institutional platforms may provide a potential infrastructure for TAI development and evaluation. For example, as participants suggested, the Mayo *Clinic*[43] *has* established a secure, subscription-based platform with anonymized patient data, offering developers access to training environments and validation pipelines across diverse patient populations.

## Discussion

Our analysis demonstrates that meta-research has the potential to be a critical and actionable pathway for addressing some of the most pressing challenges in TAI in healthcare. These challenges encompass tackling the dynamic and complex nature of TAI, integrating AI into real-world clinical environments, ensuring robustness, replicability, and reproducibility, choosing suitable evaluation metrics, advancing preclinical research, reducing the lack of transparency in the TAI field, and fostering common concepts and understanding between meta-research and TAI and beyond.

Overall, our workshop was successful in highlighting the potential contributions of meta-research to TAI, suggesting that meta-research can serve as a reflexive infrastructure supporting the development and evaluation of medical TAI. Specifically, (1) Meta-research might enable *epistemic and ethical alignment*, by surfacing, systematizing, and monitoring of AI concepts, tools, scientific practices, and the technical and ethical tensions inherent in AI development, research and evaluation. It hereby supports transparency and accountability by making research practices legible, to peers, clinicians, regulators, and patients, without prescribing and solely focusing on rigid technical solutions; and (2) Meta-research might serve as *methodological stabilization*, by promoting shared expectations for evaluation across rapidly evolving AI methods and tools. As a form of reflexive governance that supports both innovation and accountability, it might not just describe and synthesize outcomes but also shape how evidence, evaluative norms and standards are and should be established. By comparing practices across domains, meta-research can help identify structural risks that exist also in other medical domains (e.g., systemic bias, irreproducibility) rather than only identifying isolated technical failures, and systems from experimental settings into real-world clinical practice. Through these functions, (3) Meta-research might help facilitate *responsible translation*: by supporting the responsible movement of AI, meta-research has the potential to transform TAI from a fragmented and fast-moving research field into a more coherent, accountable, and clinically meaningful enterprise, providing guidance across the whole AI lifecycle. In what follows, we present a roadmap for research which is structured along the phases of the AI development lifecycle and delineates when and where meta-research can intervene, highlighting its role from development and validation through deployment and post-deployment monitoring.

## Catalogue of ideas and roadmap for future activities

We have translated the most important ideas elicited by the workshop into a catalogue of ideas for proposed meta-research activities (Table 1). In the following section we propose a roadmap based on the catalogue, focusing on the most pressing activities in our view and the stakeholders who should be involved in these activities.

| TAI Challenge | Proposed Meta-research efforts |
|---|---|
| ***A) Dynamic and Complex Nature of TAI Ethical Requirements and Principles*** | A1) Encourage structured analysis of competing principles by highlighting tensions in AI research<br>A2) Provide guidance for recognizing, managing, and resolving conflicts in medical AI studies<br>A3) Supply tools and frameworks to evaluate and ensure the trustworthiness of AI research |
| ***B) Common Terminology and Understanding of TAI*** | B1) Motivate original conceptual development grounded in diverse theoretical frameworks<br>B2) Advance transdisciplinary collaboration to build shared conceptual foundations<br>B3) Encourage synthesis of existing definitions and ethical principles to bridge epistemic gaps |
| ***C) Ensuring Robustness, Replicability, and Reproducibility*** | C1) Develop evaluation guidelines to assess robustness in AI systems<br>C2) Advocate for reproducibility during peer review, including access to data, code, and executable pipelines<br>C3) Address domain-specific challenges using lessons from other fields<br>C4) Support evaluation of existing guidelines and elevation of reporting standards to advance sustainable reproducible AI research |
| ***D) Choosing Adequate Evaluation Metrics*** | D1) Support development of standardized evaluation frameworks<br>D2) Advance efforts to align metrics with specific clinical tasks and use cases<br>D3) Improve stakeholder understanding and comparison of outcomes<br>D4) Apply existing frameworks or develop new ones to assess whether AI research studies employ appropriate evaluation metrics |
| ***E) Lack of Transparency*** | E1) Support assessment of the current state of transparency through scoping reviews, systematic reviews, and structured paper audits<br>E2) Cultivate proactive transparency practices including pre-registration of study protocols, standardized reporting guidelines, and translation into widely accessible formats<br>E3) Inform governance and accountability mechanisms<br>E4) Promote structured disclosure of system limitations and documented errors in published work |
| ***F) Advancing Preclinical Biomedical Research*** | F1) Foster creation and monitoring the use of rigorous benchmarking frameworks to enable fair comparison<br>F2) Drive adoption of software engineering standards through required reporting in biomedical studies<br>F3) Encourage transition from new model development to established benchmarking promoting robust evaluation |
| ***G) Validation in Real-World Clinical Environments*** | G1) Encourage rigorous design and reporting of validation studies<br>G2) Inform post-market monitoring best practices<br>G3) Enhance evaluation of training datasets, methodological soundness, and clinical realism in validation efforts<br>G4) Promote institutional platforms for secure testing and monitoring<br>G5) Consolidate and critically assessing the evidence base<br>G6) Identify criteria that should trigger decommissioning and the evidence base for those criteria |

*Table 1. Catalogue of Ideas, linking the six identified TAI challenges to concrete meta-research activities to address each one, as suggested by participants during the workshop.*

## Roadmap for further Meta-research Activities: from a catalogue of ideas to an executable lifecycle program to improve medical TAI

As shown, meta-research can function as a reflexive layer that monitors and evaluates AI research practices. To move from a broad catalogue of ideas to a roadmap, the key step is to decide what must be addressed first because it enables what follows, what must be stabilized next because it defines adequate evidence, and what must then be maintained because trustworthy AI is not a one-time property. Further, to increase the utility of our roadmap, we map the roadmap activities to AI lifecycle product development stages to facilitate integration into AI research and development, namely development (FORM), validation (BUILD), deployment (LAUNCH) and monitoring (MAINTENANCE)[35]. The final roadmap below therefore prioritizes a sequence of executable actions jointly addressing the currently observed lifecycle imbalance, namely the disproportionate emphasis on algorithm development over rigorous validation and clinical effectiveness studies, limited generalizability beyond development settings, rare and heterogeneously reported external validation, and a resulting misalignment between innovation incentives and clinical needs. To allow traceability, each activity is mapped to the corresponding IDs in the catalogue of ideas.

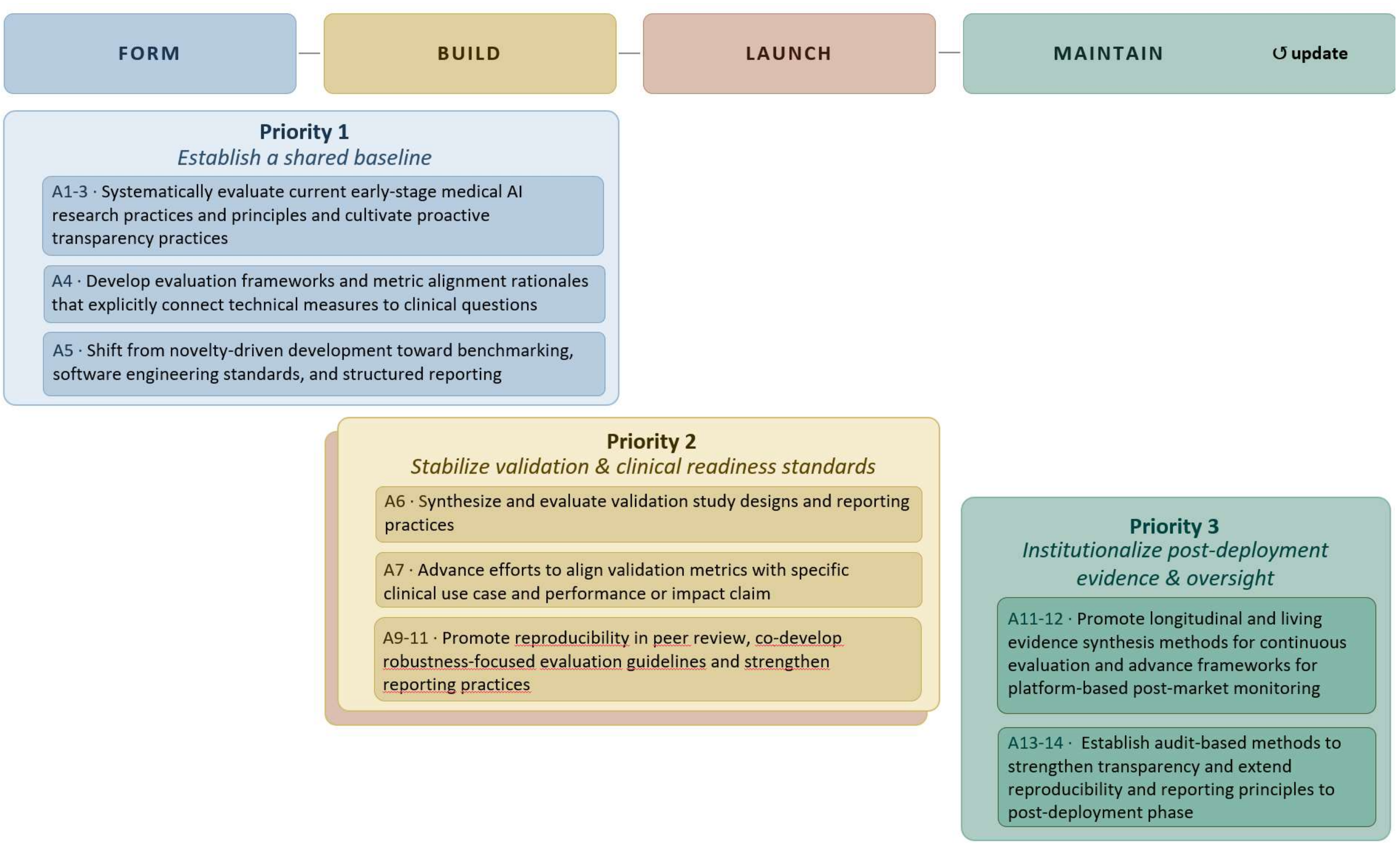

*Figure 3. Overview of the meta-research roadmap, encompassing Activities A1–A14 presented in this paper. The roadmap draws from the Catalogue of Ideas (Table 1) and organizes activities by priority across the four development lifecycle stages as proposed by Higgins and Madai (2020). Priority 2 spans both the Build and Launch phases. While each priority grouping reflects the primary lifecycle stages it targets, cross-stage relevance is acknowledged throughout.*

**Priority 1: Establish a shared evidentiary and conceptual baseline for medical TAI**

Without a shared language, shared constructs, and explicit handling of normative tradeoffs, later efforts to standardize validation, metrics, or post market monitoring will remain fragmented and contested. The goal is not to produce another list of principles, but to create an operational baseline that makes disagreements visible and manageable during early development. Concretely, this program targets the FORM stage of AI development (*What is developed, how and why?)* by making early development choices auditable, contestable, and aligned with clinical and ethical intent before technical optimization locks in assumptions.

**Activity 1** (maps to A1, A2, A3, B2, B3 in Table 1): Meta-research should begin by supporting structured analyses of competing principles and tensions in AI research, thereby directly addressing the dynamic and complex nature of TAI and providing guidance for recognizing, managing, and resolving conflicts in medical AI studies. **Stakeholders:** Meta-researchers and clinical experts should make explicit how key constructs, study designs, and modeling assumptions relate to intended clinical use, determine the comparator for impact assessment studies, and where conflicts between performance goals, clinical realism, and ethical constraints arise. Ethics boards and data stewards play a central role at this stage by contributing to the identification and management of these conflicts and by assessing whether proposed framings and data uses are compatible with acceptable use and governance requirements. **Deliverables:** A documented, structured account of the tensions, trade-offs, and justificatory choices that shape AI systems from the outset, making normative and epistemic decisions explicit rather than implicit.

**Activity 2** (maps to C4, E1in Table 1): In parallel, meta-research should support systematic assessment of current early-stage medical AI research practices through scoping reviews, systematic reviews, and structured paper audits, in line with the catalogue's focus on addressing lack of transparency. This activity aims to establish an evidence-based picture of how problem framing, dataset description, evaluation choices, and claims about clinical relevance are currently reported, and where systematic gaps or distortions occur. **Stakeholders:** Meta-researchers lead these assessments, with journals and funders as key stakeholders because they control the main enforcement points for reporting and study design expectations. Research institutions and data stewards ensure that assessments reflect realistic governance constraints. **Deliverables:** A structured mapping of the current state of transparency in early-stage medical AI research, identifying recurrent deficiencies and good practices, which can serve as a reference point for subsequent reporting and governance interventions.

**Activity 3** (maps to C2, E2, E3, E4 in Table 1): Building on these assessments, meta-research should cultivate proactive transparency practices in early AI development, including pre-registration of study protocols, specification of standardized reporting guidelines, and translation of reporting requirements into widely accessible formats. This directly operationalizes the catalogue's emphasis on improving transparency and accountability mechanisms. **Stakeholders:** Journals and funders are the primary actors for implementing these practices by embedding them in submission, review, and funding criteria, while AI researchers and clinical experts apply them in practice. Meta-researchers support the specification and refinement of these standards based on empirical findings from audits and reviews. **Deliverable:** A set of specified, field-appropriate transparency-related reporting guidelines and/or update of existing guidelines for early-stage medical AI studies, including pre-registration and standardized reporting elements that can be required and checked in subsequent project stages.

**Activity 4** (maps to D1, D2, D3, D4, G3 in Table 1**):** At the same time, and in line with the catalogue's focus on choosing adequate evaluation metrics, meta-research should support the development and refinement of standardized evaluation frameworks and advance efforts to align metrics with specific clinical tasks and use cases already in the FORM stage. The aim is to improve stakeholder understanding and comparability of outcomes and to ensure that early studies make explicit which outcomes they consider meaningful and why. **Stakeholders:** Clinicians, AI researchers, and meta-researchers jointly contribute to defining and assessing the clinical relevance of proposed metrics and evaluation approaches, while research institutions and funders provide the structural incentives to require such alignment to be made explicit in early project stages. **Deliverable:** The output is a set of early-stage evaluation frameworks and metric alignment rationales that document the link between proposed technical measures and clinical questions, establishing a reference point for later validation and benchmarking activities.

**Activity 5** (maps to C1, C3, F1, F2, F3 in Table 1): Finally, reflecting the catalogue's emphasis on advancing preclinical biomedical research, meta-research should already at this stage foster a shift away from purely novelty-driven model development toward explicit positioning of new work within emerging benchmarking cultures and toward adoption of basic software engineering standards through required reporting practices. Beyond predictive performance, new systems should specify their intended added value and how improvements in meaningful outcomes will be empirically evaluated through a comparative evaluation against the clinical status quo. **Stakeholders:** Academic medical centers and AI research institutes play a central role in shaping these expectations, journals and conferences enforce them through review criteria, and funders reinforce them through funding conditions, while industry actors contribute practical perspectives on feasibility and deployment constraints. **Deliverable:** The deliverable is the explicit embedding of comparative and methodological positioning requirements into early-stage research practice, such that new work is expected to situate itself relative to existing benchmarks and to adhere to basic software engineering and reporting standards from the outset.

**Priority 2: Stabilize evidentiary standards for validation and clinical readiness**

Building on a shared conceptual and evidentiary baseline established in the FORM stage, the next priority targets the BUILD & LAUNCH stage of AI development (*How is it validated? How is it deployed?*). In current medical AI research practice, validation and clinical evaluation are often characterized by limited generalizability beyond development settings, rare and inconsistently reported external validation, and heterogeneous validation designs and metric choices, which together undermine comparability and contribute to fragmented notions of clinical readiness[44–49]. The objective of this priority is therefore to use meta-research as a reflexive layer to clarify what counts as adequate evidence, to harmonize evaluative logics and metrics, and to stabilize validation practice in ways that support trustworthy clinical translation.

**Activity 6** (maps to D1, D2, D4, G1, G3 in Table 1**):** Meta-research should synthesize and evaluate validation study designs and reporting practices in medical AI in order to support standardized evaluation frameworks that are aligned with specific clinical tasks and use cases, and that explicitly address methodological soundness, training data evaluation, and clinical realism in validation efforts. **Stakeholders:** Meta-researchers lead the synthesis and evaluation work; clinical experts are essential for judging clinical realism and task alignment; AI researchers contribute methodological feasibility and technical constraints. Regulators and health technology assessment bodies are key stakeholders because stabilized validation frameworks support regulatory-grade and decision-relevant evidence expectations, and journals are key stakeholders because publication norms strongly shape validation reporting practice. **Deliverable:** A set of standardized, task-oriented validation evaluation frameworks, including criteria for validation design and reporting that explicitly incorporate clinical realism and methodological soundness, to serve as a shared reference point for assessing clinical readiness.

**Activity 7** (maps to D1, D2, D3, D4 in Table 1): Meta-research should advance efforts to align evaluation metrics with specific clinical tasks and use cases and improve stakeholder understanding and comparison of outcomes by assessing whether medical AI studies employ appropriate evaluation metrics, using existing frameworks where applicable and developing new ones where needed. **Stakeholders:** Clinicians, AI researchers, and meta-researchers jointly define what "appropriate" means in context and assess whether reported metrics support clinically meaningful interpretation; journals and conferences operationalize these expectations by requiring explicit justification and comparability of metric choices; funders reinforce uptake by embedding such expectations in funding criteria. **Deliverable:** A set of metric-alignment outputs that enable comparability across studies, consisting of standardized metric selection guidance tied to clinical tasks and an assessment approach for identifying misalignment between reported metrics and intended use.

**Activity 8** (maps to C1, C3, C4 in Table 1): Meta-research should develop evaluation guidelines to assess robustness in AI systems, address domain-specific challenges using lessons from other fields, and support evaluation of existing guidelines and elevation of reporting standards to advance sustainable reproducible AI research, specifically as components of validation and clinical readiness assessment rather than as optional add-ons. **Stakeholders:** Meta-researchers and methodologists develop and evaluate robustness-oriented guidance; AI researchers contribute technical knowledge about robustness testing and domain constraints; clinical experts contribute clinically relevant variability and failure considerations; journals and standards communities are key for disseminating and normalizing reporting standards. **Deliverable:** Robustness-oriented evaluation guidance for validation, coupled with strengthened reporting standards that support sustainable reproducible AI research and make robustness-related evidence interpretable and comparable across studies.

**Activity 9** (maps to C2, E2, E4 in Table 1): Meta-research should advocate for reproducibility during peer review, including access to data, code, and executable pipelines, and promote structured disclosure of system limitations and documented errors in published validation work, thereby strengthening the interpretability and credibility of validation evidence. **Stakeholders:** Journals and peer reviewers are the primary enforcement points; funders are key leverage actors for requiring reproducibility conditions; data stewards support compliant access and governance arrangements; AI and clinical researchers operationalize these requirements in validation studies. **Deliverable:** Reproducibility and disclosure requirements embedded in validation and review processes, including explicit expectations for accessible artifacts and structured reporting of limitations and errors.

**Activity 10** (maps to F1, F2, F3, G1 in Table 1): Meta-research should foster creation and monitored use of rigorous benchmarking frameworks to enable fair comparison, drive adoption of software engineering standards through required reporting in biomedical studies, and encourage transition from new model development toward established benchmarking that promotes robust evaluation, thereby strengthening the rigor and comparability of validation studies and their reporting. **Stakeholders:** Academic medical centers and AI research institutes support benchmarking and methodological infrastructure; journals and conferences enforce reporting and benchmarking expectations through review criteria; funders reinforce adoption via conditions and incentives; industry contributes perspectives on feasibility and real-world constraints relevant to robust evaluation. **Deliverable:** Validation-relevant benchmarking and reporting expectations that support fair comparison and robust evaluation, including adoption of software engineering standards through reporting requirements and explicit incentives to prioritize benchmarking-aligned evaluation over novelty-driven validation.

**Priority 3: Institutionalize post-deployment evidence generation and performance oversight**

Even when systems have passed initial validation, trustworthy medical AI cannot be treated as a static achievement. Performance, generalizability, and safety can change across time, settings, and populations (model drift), and current practice still relies too heavily on one-off, static evaluations. This priority therefore targets the MAINTENANCE stage of AI development (*How is performance stability ensured?*) and meta-research enables continuous, post-deployment evidence generation, synthesis, and oversight facilitating responsible translation.

**Activity 11** (maps to C4, G2 in Table 1): Meta-research should structure post-deployment evidence generation through longitudinal reviews, registries, and living syntheses, rather than relying on single, static evaluations, to support systematic oversight of real-world performance, generalizability, and model drift. **Stakeholders:** Meta-researchers design and maintain longitudinal review and synthesis approaches; healthcare providers and health systems contribute real-world performance data and clinical context; regulators and health technology assessment bodies use these evidence streams to inform oversight and decision-making. **Deliverable:** A set of post-deployment evidence synthesis mechanisms, including longitudinal review protocols, registry-based monitoring approaches, and living synthesis methods, that enable continuous evaluation of AI systems across time, settings, and populations.

**Activity 12** (maps to G2, G4 in Table 1): Meta-research should promote the development and use of institutional platforms for secure testing and monitoring and inform best practices in post-market monitoring, so that real-world performance assessment can be carried out in a controlled, auditable, and methodologically sound manner. **Stakeholders:** Healthcare institutions and academic medical centers host and operate secure testing and monitoring platforms; methodologists specializing in AI impact assessment regulators and post-market surveillance authorities define oversight expectations; AI developers and industry actors contribute systems and technical interfaces for monitoring; data stewards ensure governance-compliant data access and use. **Deliverable:** Operational post-market monitoring infrastructures and platform-based testing environments, together with documented best-practice guidance for their use in continuous performance assessment.

**Activity 13** (maps to: E1, E3, E4 in Table 1): Meta-research should support systematic assessment of transparency in post-deployment evaluation through structured audits and reviews and promote structured disclosure of system limitations and documented errors in real-world use, thereby strengthening governance and accountability mechanisms beyond initial approval or publication. **Stakeholders:** Meta-researchers conduct audits and reviews of post-deployment reporting and disclosure practices; journals, regulators, and oversight bodies act as enforcement points for disclosure and accountability expectations; healthcare providers and developers contribute incident reports, limitations, and error documentation. **Deliverable:** A structured, post-deployment transparency and accountability framework, including audit-based assessments and standardized disclosure practices for limitations and errors observed in real-world use.

**Activity 14** (maps to C2, E2 in Table 1): To ensure that post-deployment evidence remains interpretable and trustworthy, meta-research should extend reproducibility and transparency requirements into the maintenance phase, including continued access to relevant data, code, and executable pipelines where feasible, and the use of standardized reporting formats for post-market evidence generation. **Stakeholders:** Journals, regulators, and oversight bodies define and enforce post-deployment reporting and reproducibility expectations; data stewards support compliant access arrangements; AI developers and healthcare institutions implement these requirements in ongoing monitoring and update cycles. **Deliverable:** Reproducibility and transparency requirements embedded in post-deployment evaluation and reporting processes, ensuring that real-world performance evidence can be scrutinized, compared, and reused for longitudinal assessment and regulatory oversight.

Building on the points above, our roadmap can guide TAI toward genuine practical impact along the *full* AI lifecycle.

## Focused Observations Derived from the Workshop Material

In the last section of our paper, we would like to highlight two important observations and recommendations which surfaced throughout the workshop.

### *Clarifying the Distinction Between the Tool Level and the Study Level Robustness and Transparency*

One of the main points brought up by the participants was confusion around the meaning of robustness. This confusion may be reduced by more clearly distinguishing between the tool level and the study level, particularly for central concepts, such as robustness and transparency: tool-level robustness refers to the stability and reliability of the AI system across varying inputs and contexts[23], while study-level robustness concerns the methodological soundness of the research design[50]. While not the same, they are intertwined, as robustness of a tool can only be proven by a robust study. Similarly, transparency of the tool differs from study transparency: while tool-level transparency, such as revealing model architecture, training data provenance, and version history[21], does not necessarily require publication within a formal study (though doing so can greatly enhance overall transparency), study-level transparency specifically entails the clear documentation of study conduct, like protocols, evaluation metrics, sample sizes and eligibility criteria[51]. While the two dimensions differ, they are closely interconnected and mutually reinforcing. Conceptual confusion can, in the short term, be minimized by clearly distinguishing between the two levels, and, in the long term, by developing concepts that span both. To our knowledge, there is only one definition of robustness translating a classic meta-research definition of robustness (the combination of reliability and validity) to an integrative meta-research and TAI definition[52]. Understanding and addressing both dimensions – the tool and the related studies – is essential for building trust in AI systems, as successful validation needs to encompass robust and transparent tools *and* robust and transparent research.

### *The Analogy Between AI and Pharmaceutical Development*

Another notable observation concerned participants' frequent use of the analogy between AI and pharmaceutical development when reflecting on challenges and meta-research approaches. This analogy is useful in highlighting the need for rigorous validation, effectiveness assessment, post-market surveillance, and transparency throughout the lifecycle of AI systems[35], mirroring the clinical trial phases, regulatory oversight, and pharmacovigilance in drug development[53–56]. However, the analogy should not be extended uncritically. Unlike drugs, which are typically characterized by stable and well-defined chemical compositions, AI systems are socio-technical artefacts that are dynamic and context-sensitive. They may change over time through software updates, retraining, shifts in data distributions, or adaptations within the clinical environment.

In this sense, the intervention itself is not necessarily fixed. Importantly, this does not mean that AI systems cannot be evaluated using causal impact assessment. As in pharmacology, effectiveness can be established without full knowledge of the underlying mechanism or "active ingredient." Rather, the challenge lies in the structural features of AI systems, their instability over time, update dynamics, and embedding within complex sociotechnical settings, which complicate the design and interpretation of impact evaluations. Meta-research and TAI can therefore benefit from the pharmaceutical analogy as a heuristic, while remaining attentive to the dynamic and context-dependent nature of AI interventions.

### Limitations and Strengths

Our study is based on an interdisciplinary workshop followed by a thematic analysis of the materials produced by participants. While this approach offers rich insights into how researchers from TAI and meta-research conceptualize the intersection of their fields, several limitations must be acknowledged. First, the workshop format likely introduced selection bias, as participants were already engaged or interested in questions of research integrity and AI ethics and are no representative sample of the scientific community. Second, while the workshop aimed to promote inclusivity, researchers with certain attributes, such as a strong inclination toward writing, may have contributed more prominently or unintentionally overshadowed other voices. Third, not all workshop discussions were documented in writing. As documentation was produced by subgroups during exercises, some points remained undocumented when groups chose not to record them. The synthesis therefore reflects the recorded output rather than the full set of perspectives exchanged during the workshop. Fourth, the thematic analysis methodology is inherently interpretive and may reflect framing decisions made by the researchers, despite efforts to ensure rigor and reflexivity. This also means that our roadmap should be understood as the first step towards a normative research agenda rather than a conclusive final framework.

At the same time, the study offers several strengths. By bringing together, for the first time, experts from meta-research and TAI, the workshop facilitated cross-field reflection that would be difficult to achieve through literature review or survey methods alone. The workshop format employed in this study demonstrated strong potential for interdisciplinary knowledge generation. Facilitated group work and collaborative writing exercises enabled the identification of nuanced, cross-cutting themes and structured yet flexible exploration of shared challenges across disciplinary boundaries. The in-person setting fostered rich discussion, immediate feedback, and iterative refinement of ideas. Moreover, the integration of design thinking principles and digital collaboration tools supported inclusive participation and sustained focus throughout the sessions. These enabled the identification of nuanced, cross-cutting themes. The analysis also moved beyond abstract principles by identifying specific, actionable connections between TAI challenges and meta-research practices, particularly in relation to robustness and transparency. This paper contributes not only to clarifying the conceptual relationship between TAI and meta-research but also offers direction for future research by showing how meta-research can strengthen the scientific foundations of medical AI. In doing so, it supports the development of AI tools that are grounded in rigorous, evidence-based practices — ultimately advancing the goal of genuinely trustworthy AI.

## Conclusion

Meta-research has the potential to be a critical and actionable pathway for addressing several of the most pressing challenges in medical TAI. Our workshop was successful in eliciting a rich catalogue of ideas for further research to integrate meta-research activities in TAI. Our resulting roadmap can serve as the first normative step to translate these activities into research priorities for involved stakeholders.

**Acknowledgments** We would like to acknowledge the professional and friendly organization of the workshop by the staff of the Volkswagen Foundation. We also would like to acknowledge the professional facilitation by Prof. Dr. Anthony Giannoumis, whose presence, professionalism, and humor were the foundation of our successful workshop.

Timo Minssen's contribution to this article was further supported by the Novo Nordisk Foundation (NNF) through a grant for the scientifically independent Collaborative Research Program in Bioscience Innovation Law (Inter-CeBIL Program – Grant No. NNF23SA0087056).

**Author contributions statement** Conceptualisation: V.B., M.B., J.F., R.M., T.W and V.I.M. Formal analysis: V.B., E.M. and I.V. Writing – original draft: V.B., M.B., J.F., R.M., E.M., I.V. and V.I.M. Writing – review and editing: M.B., J.F., R.M., E.M., I.V., T.W., L.A.M.S., J.A., A.B.B., D.B.B., K.H., B.H., M.H., E.H., S.H., A.d.H., S.K., S.M., T.M., M.S.N., N.P., J.L.R., T.R.H., M.S., K.T., E.V. and M.W. Visualisation: V.B. Supervision: T.W. and V.I.M. Project administration: V.I.M. Funding acquisition: V.I.M.

All authors read and approved the final manuscript.

**Competing interests statement** All authors declare no financial or non-financial competing interests. The Volkswagen Foundation funded the workshop logistics and participant travel but had no role in the study design, data collection, analysis, interpretation, or the decision to submit for publication. The Novo Nordisk Foundation grant supporting T.M. similarly had no role in the design or conduct of this study.

**Data availability statement** The materials collaboratively generated during the workshop and analysed in this study are not publicly available as participants did not consent to public release of the raw data. Anonymised outputs are available from the corresponding author upon reasonable request.

## Appendix

| Number | Gender (m/f/n) | Position | Field of Research |
|---|---|---|---|
| 1 | f | Assistant professor | Meta-research |
| 2 | f | Senior Researcher | Trustworthy AI, Ethics, Policy |
| 3 | f | Researcher | Trustworthy AI, Meta-research |
| 4 | f | PI, statistics and epidemiology | Meta-research, Epidemiology |
| 5 | m | Postdoctoral fellow | Meta-research |
| 6 | m | Team lead, PI | Trustworthy AI, Meta-research |
| 7 | f | AI researcher and consultant | Trustworthy AI, Policy and Law |
| 8 | f | PhD candidate | Meta-research, Validation of AI |
| 9 | m | Director of Research | Trustworthy AI, Ethics |
| 10 | m | Professor of Ethics | Trustworthy AI, Medical Ethics, Neuroethics |
| 11 | f | Professor for Meta-research | Meta-research |
| 12 | m | Group lead | Meta-research, Computer Science |
| 13 | m | Professor of statistics | Meta-research, Validation of AI |
| 14 | m | Professor of computer science | Trustworthy AI, Computer Science |
| 15 | m | Associate professor | Trustworthy AI, Law |
| 16 | f | Full professor | Trustworthy AI, Ethics, Computer Science |
| 17 | m | Associate professor | Trustworthy AI, Applied Ethics |
| 18 | m | Professor of Law | Trustworthy AI, Law |
| 19 | f | Lead Strategy and Innovation | Trustworthy AI, Ethics, Policy |
| 20 | f | Lead, PI | Trustworthy AI, Meta-research |
| 21 | f | Professor of economics | Trustworthy AI, Ethics Operationalization |
| 22 | f | Research fellow | Trustworthy AI, Meta-Research |
| 23 | f | Postdoctoral fellow | Trustworthy AI, Meta-Research |
| 24 | m | Professor of statistics | Meta-research, Computer science |
| 25 | f | Postdoctoral fellow | Meta-research |
| 26 | f | Professor | Meta-research |
| 27 | m | Senior researcher | Meta-research |
| 28 | m | Researcher | Meta-research |
| 29 | f | Senior Researcher | Meta-research |

**Table A1.** Participant characteristics. All information is based on self-reported data collected through the pre-workshop questionnaire.